\documentclass[fleqn,usenatbib]{mnras}
\usepackage{newtxtext,newtxmath}
\usepackage[T1]{fontenc}

\DeclareRobustCommand{\VAN}[3]{#2}
\let\VANthebibliography\thebibliography
\def\thebibliography{\DeclareRobustCommand{\VAN}[3]{##3}\VANthebibliography}

\newcommand{\bicho}{V4641\,Sgr}
\usepackage{graphicx}	% Including figure files
\usepackage{amsmath}	% Advanced maths commands

\title[Radio jets in V4641 Sgr aligned with its UHE bubble]{Alignment of radio jets in the microquasar V4641 Sagittarii with its high-energy structures}

\author[J. Mart\'i and P.L. Luque-Escamilla]{
Josep Mart\'i,$^{1}$\thanks{E-mail: jmarti@ujaen.es}
Pedro L. Luque-Escamilla,$^{2}$
\\
$^{1}$Departamento de F\'isica. Universidad de Jaén, EPS Jaén. Campus Las Lagunillas s/n, Ja\'en, 23071, Ja\'en, Spain\\
$^{2}$Departmento de Ingeniería Mecánica y Minera, Universidad de Jaén, Campus Las Lagunillas s/n, Ja\'en, 23071, Ja\'en, Spain
}

\date{Accepted 2025 November 22. Received 2025 November 22; in original form 2025 October 24}

\pubyear{\the\year{}}

\begin{document}
\label{firstpage}
\pagerange{\pageref{firstpage}--\pageref{lastpage}}
\maketitle

\begin{abstract}
V4641 Sagittarii (\bicho) is a unique Galactic microquasar system featuring a stellar-mass black hole accreting matter from a massive companion. One of its intriguing features is the presence of relativistic radio jets almost perpendicular to  the observed extended gamma-ray emission, implying significant propagation effects or interactions with the Galactic magnetic field. Here we report observational evidence that the radio jet and the very high-energy (VHE) and ultra high-energy (UHE) gamma-ray emission could be aligned along a common axis, indicating a co-spatial or co-directional origin. This alignment supports a model where  synchrotron radio emission, VHE and UHE gamma rays are produced within a single, highly collimated relativistic outflow. Our findings favor scenarios of in-situ particle acceleration up to hundreds of TeV, challenge previous interpretations involving large-scale particle diffusion, and simplify the geometric modeling of the source. This case highlights the potential of \bicho\ as a PeVatron candidate within our Galaxy and provides a benchmark for understanding jet composition and magnetic structure in microquasars.
\end{abstract}


%%%%%%%%%%%%%%%%%%%%%%%%%%%%%%%%%%%%%%%%%%%%%%%%%%

%%%%%%%%%%%%%%%%% BODY OF PAPER %%%%%%%%%%%%%%%%%%

\section{Introduction}

Relativistic jets launched from compact objects are among the most energetic and enigmatic phenomena in the Universe. In the Galactic environment, the so-called microquasars, which are X-ray binaries featuring a stellar-mass black hole or neutron star accreting from a companion, serve as nearby analogues to active galactic nuclei, allowing high-resolution studies of jet formation, propagation and interaction with the interstellar medium. 
Among them, \bicho\ (also known as SAX J1819.3–2525) has long stood out for its extreme variability, powerful outbursts, and complex jet behavior. 
First discovered in 1999 as an X-ray source, it was proposed as a microquasar later that same year, when an extraordinary radio outburst was observed with the Very Large Array (VLA) of the U.S. National Radio Astronomy Observatory (NRAO), 
which revealed superluminal jet motions \citep{Hjellming_2000}. 
While these authors inferred a jet inclination of about 63$^{\circ}$ to the line of sight, subsequent optical modeling by \citet{Orosz2001} constrained the inclination to $\le 12^{\circ}$, indicating that the jet is in fact oriented almost perpendicular to the plane of the sky.
The emission seemed coincident with a B-type subgiant star orbiting a dynamically confirmed black hole with an estimated mass of $\sim 6$ M$_{\odot}$  and with orbital period of $2.82$ days \citep{MacDonald2014, Orosz2001}.
The stellar companion in the \bicho\ has a \textit{Gaia} DR3 parallax  of $0.1692\pm0.0262$ mas, which translates to $ d = 5.9\pm0.8$ kpc \citep{gaia2022}.
Since then, new outbursts of \bicho\ have occurred 
with a median recurrence time of $\sim 220$ days \citep{Tetarenko_2016}, but always much less intense than those in 1999, which has prevented the obtention of a new radio map to confirm the proposed jet structure.

In recent years, \bicho\ has emerged as one of the most compelling Galactic laboratories for particle acceleration at both high and very high energies. Observations with 
High Altitude Water Cherenkov  Observatory (HAWC) and Large High Altitude Air Shower Observatory (LHAASO) have firmly established VHE and UHE gamma‐ray emission from an extended region with a jet-like structure almost in the plane of the sky (ranging from TeV  to beyond 100 TeV), with detections in UHE up to $\sim 0.8$ PeV with a remarkably hard spectrum, making it a confirmed PeVatron microquasar \citep{Alfaro_2024,lhaaso24}.
The High Energy Stereoscopic System (H.E.S.S.) collaboration also detected the source in the TeV range \citep{HESS2025}.
In September 2024, Xtend instrument onboard X-Ray Imaging and Spectroscopy Mission (XRISM) unveiled extended X‑ray emission out to $~7$ arcmin ($\sim 13$ pc), indicating synchrotron or thermal components from particles accelerated within $\sim 10$ pc of the black hole, and revealing insights into ambient magnetic fields ($\sim 80\, \mu$G) and diffusion suppression at $\sim 100$ TeV \citep{Suzuki2025}. Complementary studies propose that cosmic rays escape along Galactic magnetic field lines, creating a $\sim 200$ pc gamma‑ray jet–like structure and suggesting detectability of neutrinos in forthcoming km$^3$ detectors \citep{Neronov2024}. Together, these multi‑messenger discoveries position \bicho\ as a cornerstone for understanding microquasar‐driven cosmic‑ray acceleration and the role of stellar‐mass black holes in Galactic high‑energy phenomena.

Among the many intriguing properties that make this system unique, one of the most striking is that the large-scale gamma-ray emission
appears strongly misaligned with a relative inclination of approximately $60^\circ$--$72^\circ$ \citep{Alfaro_2024} 
with respect to the relativistic radio jet \citep{Hjellming_2000,Orosz2001, Chaty2003,Gallo2014}.
Although the jet power could be energetically capable to maintain the observed gamma-ray emission \citep{Alfaro_2024},
this configuration prevents the natural injection of the energetic particles responsible for the gamma-ray emission, which seem to be hadrons \citep{Alfaro_2024,lhaaso24}, from the relativistic jet of the microquasar where they are accelerated. This challenging issue has led to the development of more sophisticated models, such as the one proposed by \citet{Neronov2024} in which anisotropic diffusion of accelerated hadrons escaping from the jet, guided by the large-scale magnetic field of the Galaxy, are responsible for the observed bipolar morphology. 
Although alternative mechanisms could be considered to account for the high-energy emission of \bicho, they would necessarily involve more complex and exotic physical scenarios.

Here, we present a new analysis of the key historical VLA  observations
of \bicho\ that provides strong evidence for a spatial alignment between the radio jet and the axis of VHE gamma-ray emission. 
In line with the principle of Occam’s razor, our interpretation of the archival VLA data  
questions the need to invoke large-scale particle diffusion or magnetic field deflection. Instead, the results support a scenario in which both emissions arise from a co-spatial relativistic outflow. If confirmed, this finding would have significant implications for our understanding of particle acceleration in Galactic jets, positioning V4641 Sgr as a benchmark system for testing models of jet physics in compact objects.

\section{Data analysis and results}

\begin{table}
\centering
\caption{Summary of VLA radio observations of \bicho\ reported by \citet{Hjellming_2000} used in this work.}
\label{tab:vla_obs}
\begin{tabular}{lccr}
\hline
Date (UT, 1999) & MJD & Freq. (GHz) & Flux Density (mJy)\\
\hline
Sep 16.027 & 51437.027 & 4.9  & $420 \pm 20$  \\
Sep 16.048 & 51437.048 & 4.9  & $400 \pm 20$  \\
\hline
\end{tabular}
\end{table}

\begin{figure*}
\includegraphics[width=.53\linewidth]{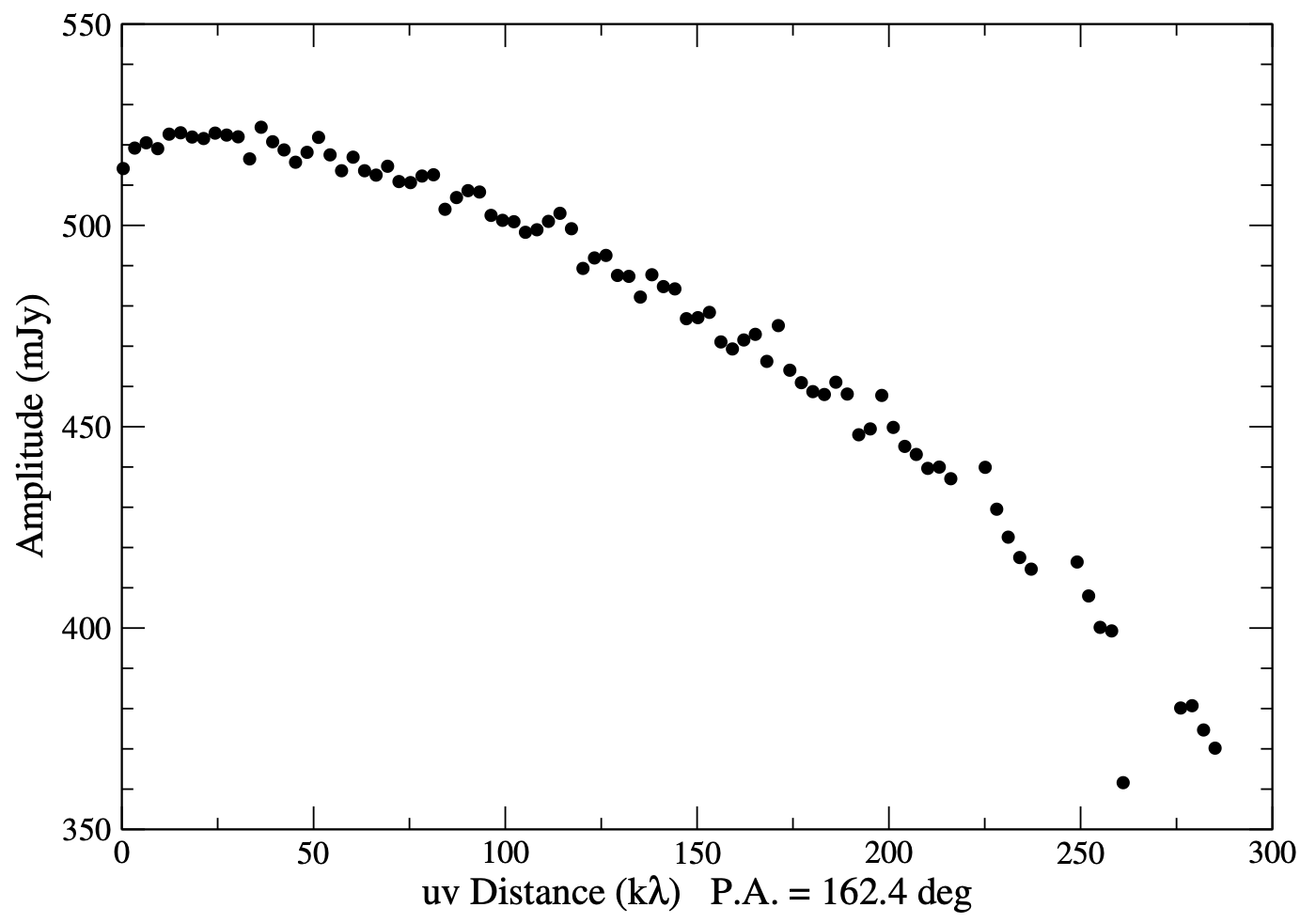}\hfill
\includegraphics[width=.45\linewidth]{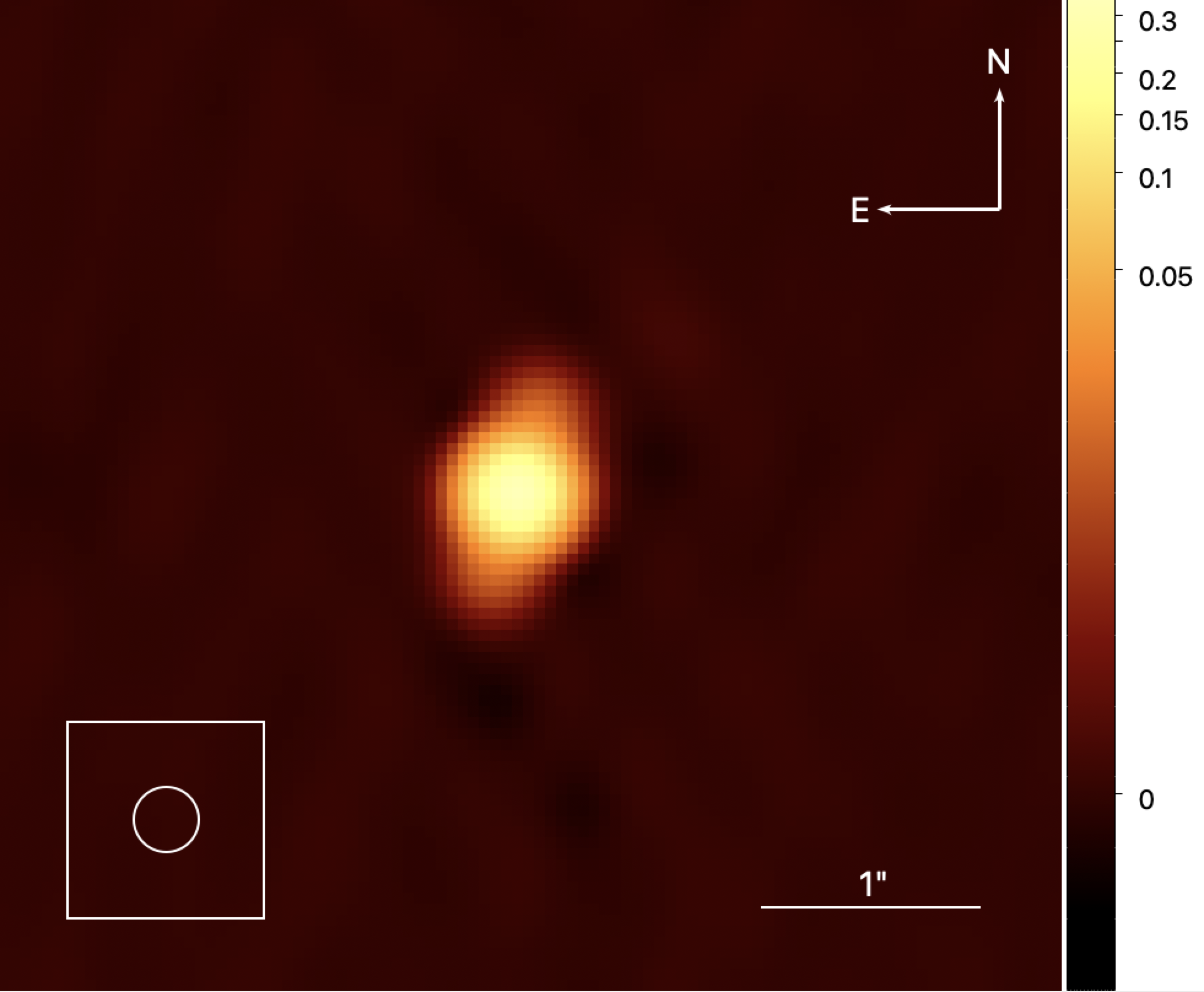}
 \caption{\textit{Left:} Visibilities as a function of baseline length at a position angle of $162^\circ$, using a total of 100 bins. Error bars are smaller than the symbol size.
\textit{Right:} Over-resolved map of \bicho\ showing diffuse extended emission. The restoring beam is circular with a FWHM of $0.3^{\prime\prime}$. The horizontal bar indicates a scale of $1^\prime$. Colour bar on the right gives the flux density scale in units of Jy beam$^{-1}$, with the rms noise of the map
being 0.15 mJy beam$^{-1}$. North is up, and East is to the left.} \label{fig:uvplot}
 \end{figure*}

\begin{figure}
\includegraphics[width=0.90\linewidth]{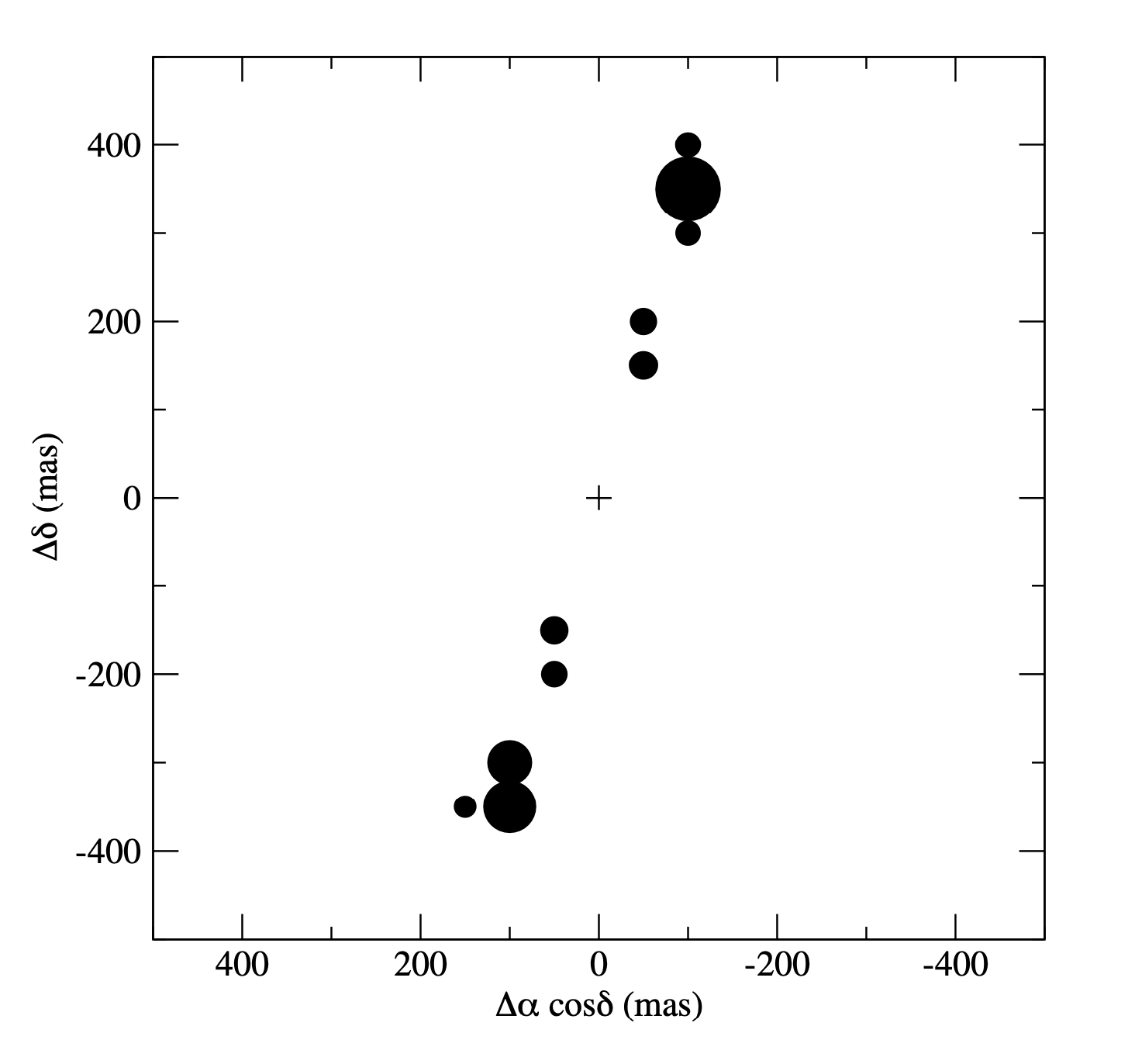}
\caption{Zoom-in on the \textsc{\textsc{\textsc{clean}}} components of the over-resolved radio map shown in Fig.~\ref{fig:uvplot}, revealing an alignment suggestive of a bipolar, elongated structure. The area of each circle is proportional to the flux density of the corresponding \textsc{\textsc{\textsc{clean}}} component. The small cross marks the position of the central core component, which is too bright to be represented using the same brightness scale.}\label{fig:clc}
\end{figure}

Radio observations of \bicho\ are very limited. Apart from the campaign by \citet{Hjellming_2000} during the 1999 outburst, no VLBI follow-up or significant radio coverage has been reported for later events. In their seminal study these authors presented a heterogeneous radio data set collected with seven different facilities (GBI, VLA, MOST, ATCA, OVRO, RATAN-600, and MERLIN). Rather than following a uniform reduction strategy, they adopted a strongly model-driven approach. For the earliest VLA observations at 4.9 GHz (September 16.027 and 16.048 UT),
during the maximum of radio emission,
 visibilities were fitted with simple geometric models such as elliptical Gaussians and combinations of fixed point sources. After several trials, they converged on a three-component model (two fixed points plus an elongated Gaussian), which was used to argue for the presence of a highly relativistic jet. Imaging was also carried out using the Non-Negative Least Squares (NNLS) algorithm, which is less commonly applied in radio interferometry but was preferred by the authors over \textsc{\textsc{\textsc{clean}}} for sources deemed marginally resolved. The resulting maps showed an elongation roughly aligned with the synthesized beam, and were interpreted as evidence for superluminal ejection of jets with a position angle P.A. $\simeq 162^\circ$.

This methodology, while innovative, is unconventional and introduces significant model dependence. The early adoption of restrictive geometric fits 
could bias 
the analysis towards elongated structures, while the reliance on NNLS makes the results harder to reproduce or compare with later studies that rely on standard deconvolution techniques. Moreover, \citet{Hjellming_2000} combined flux measurements from multiple telescopes and frequencies into global light curves that were then fitted with simplified “finite jet segment” models. Such heterogeneous data, with varying calibrations and sensitivities, can produce systematic inconsistencies that are difficult to disentangle from genuine source behavior. In this sense, their treatment represents less a neutral reduction of the interferometric data and more a physically motivated reconstruction consistent with the relativistic jet hypothesis.

We therefore re-analyze the same most important VLA datasets from 1999 (see Table~\ref{tab:vla_obs}) in order to independently verify the published results.
 Our data reduction was carried out using the AIPS software package of NRAO (31DEC24 version). 
The original VLA  run did not include an amplitude calibrator and only the phase calibrator 1744-312 was observed, for which we assumed a flux density of 0.42 Jy\footnote{https://science.nrao.edu/facilities/vla/observing/callist}.
The two data blocks were iteratively self-calibrated in phase starting with an initial point source model.
Separate imaging rendered comparable results so that we merged them
into a single dataset as flux density variability was moderate (not exceeding 5\%). In the left panel of Fig. \ref{fig:uvplot}), we show 
the correlated amplitudes from individual antenna pairs that essentially reproduce the visibility analysis presented by \citet{Hjellming_2000}.
We binned data into 
100 intervals of projected baseline length along 
the same P.A. proposed
by these authors. As they already found, it is clear that
the source was resolved less than a day after the maximum of the X-ray outburst given the clear amplitude decay as the baseline lengths increase.

To produce our radio maps, we deliberately adopted a simple and transparent reduction procedure based on \textsc{\textsc{clean}}, the standard deconvolution algorithm in radio astronomy, and avoided unnecessary manipulation of the data. We believe this approach enhances the reliability of the resulting images, as we justify below.

\citet{Hjellming_2000} cautiously argued against the use of \textsc{clean} on their 1999 September 16 data, since the observed extension, roughly along the North–South direction, was closely aligned with the major axis of the synthesized beam.
These authors also expressed concern that the source size, approximately $ 0.25''$ and therefore smaller than the synthesized beam of $\sim 0.8^{\prime\prime} \times 0.3^{\prime\prime}$, using uniform weighting of visibilities. Under such conditions, the application of the \textsc{\textsc{\textsc{clean}}} algorithm was considered potentially problematic due to the risk of producing artifacts.
However, our results yield a source size that is approximately twice as large as that reported by these authors (see right pannel in Fig.~\ref{fig:uvplot}), and with a P.A. $\simeq 162^\circ$, which differs significantly from the  P.A. $\simeq 169^\circ$ of the synthesized beam. Therefore, the cautious concerns raised by \citet{Hjellming_2000} could be not so relevant.

Moreover, it is now well established that when the signal-to-noise ratio (SNR) is sufficiently high, 
it is possible to restore the \textsc{\textsc{clean}} components with a certain degree of over-resolution, thereby recovering spatial information that remains encoded in their distribution \citep{valencianos}. 
This is precisely the situation encountered for \bicho\ on 1999 September 16, when the source was nearly as bright as the phase calibrator. 
In fact, the averaged visibility amplitudes shown in the left panel of Fig.\ref{fig:uvplot} and the corresponding image-plane data reach SNR values as high as $\sim10^3$. 
Therefore, why not take advantage of this circumstance and legitimately apply some degree of over-resolution to \bicho?
According to \citet{valencianos}, the minimum detectable size $\theta_M$ of a radio source, in terms of the full width at half maximum (FWHM) of the restoring beam,
 is given by:
\begin{equation}
\theta_M = \beta  \left(\frac{\lambda_c}{2 \,\rm{SNR}^2} \right)^{1/4} \times \rm{FWHM},
\end{equation}
where $\beta$ and $\lambda_c$ are slowly varying functions, typically in the ranges $\beta \simeq 0.5$–$1.0$ and $\lambda_c \simeq 3.8$–$8.8$, respectively. For our data, this relation suggests that structures down to $\sim$10\% of the nominal FWHM could, in principle, be detected.

Nevertheless, to remain conservative and minimize potential artifacts, we limit the over-resolution to a factor of two. Accordingly, we adopt a circular restoring beam of $0.3^{\prime\prime}$
that is about half the nominal FWHM 
with uniform weight 
to produce  
the map in the right panel in Fig. \ref{fig:uvplot} 
A clearly symmetric and elongated morphology emerges from this modest application of over-resolution.
In Fig.~\ref{fig:clc}, we present an enlarged view of the \textsc{\textsc{\textsc{clean}}} components that support our interpretation of a symmetric bipolar outflow from \bicho. The radio emission appears elongated along P.A. $=163 \pm 1^\circ$, which is in accordance to the value obtained by \citet{Hjellming_2000}. This is close to ---but significantly different from--- the P.A. $= 169^\circ$ of the synthesized beam. With appropriate caution, we interpret this $\simeq 6^\circ$ offset as evidence that the observed bipolar structure is intrinsic to the source and not an artifact introduced by the imaging process.

\begin{figure}
\includegraphics[width=1\linewidth]{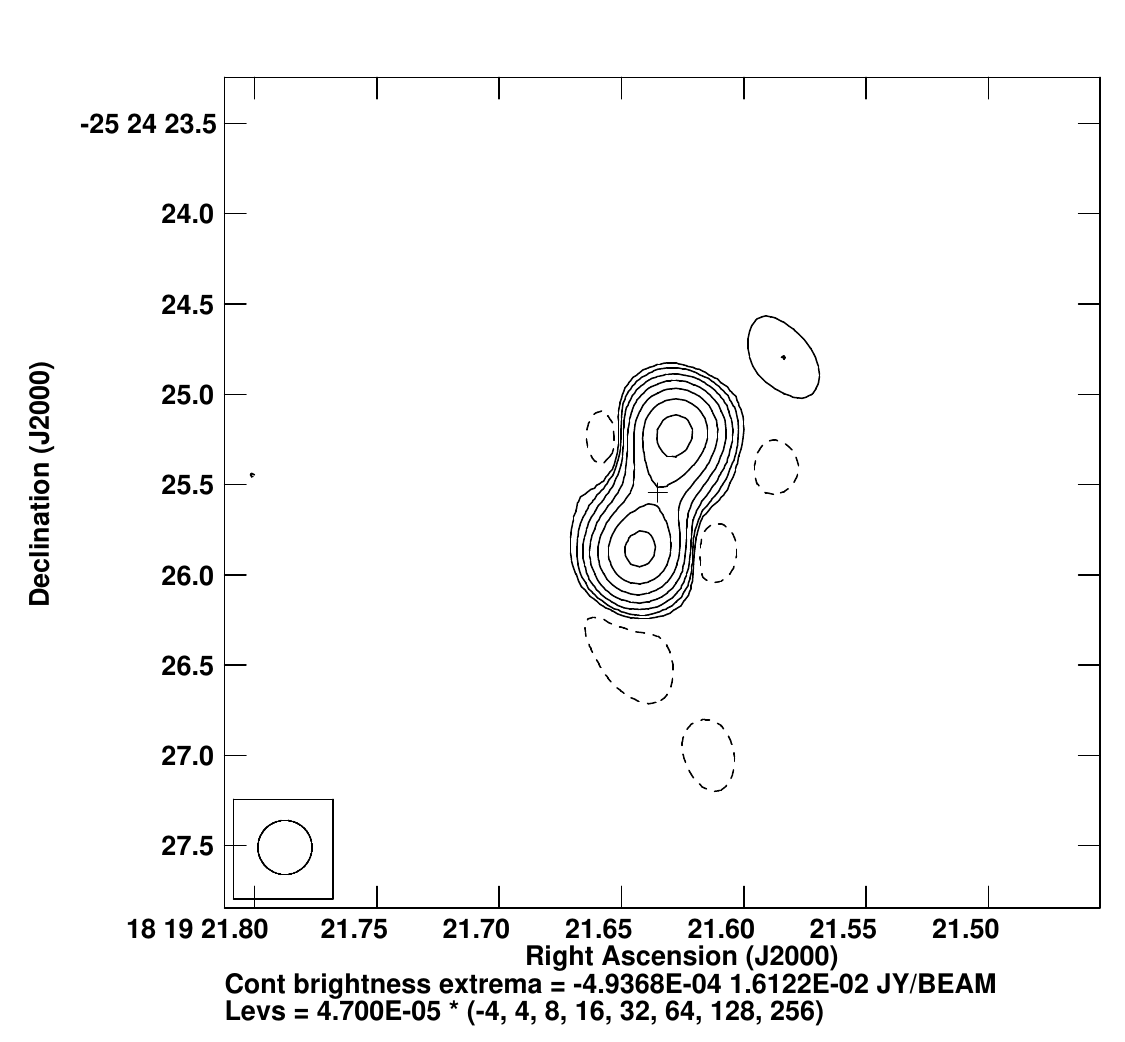}\hfill
\caption{The same over-resolved map of V4641 Sgr as in Fig.~\ref{fig:uvplot} but with
the central \textsc{clean} component subtracted at the cross position as described in
the text. The circular synthesized beam shown in the bottom right corner also
has a FWHM of $0.3^{\prime\prime}$. 
}\label{fig:blocks}
\end{figure}

In an effort to enhance the visibility of the extended emission in the over-resolved map shown in Fig.~\ref{fig:uvplot}, we attempted to suppress the compact core by subtracting the central \textsc{clean} component, which accounts for more than 90\% of the total flux density, and then re-imaged the residuals
The resulting map is displayed in Fig.~\ref{fig:blocks}. This subtraction was carried out to provide estimates of the flux densities of the northern and southern lobes, as well as their angular offsets with respect to the ejection site. These measurements are summarized in  Table~\ref{tab:prop} and are discussed below.

\begin{table}
\centering
\caption{\bicho\ parameters obtained from the substracted radio map. }
\label{tab:prop}
\begin{tabular}{lc}
\hline
Parameter [Units] & Value   \\
\hline
North lobe angular offset [mas] & $\Delta \alpha \cos \delta = -93\pm 10$\\
        & $\Delta \delta = +321\pm 10$\\
South lobe angular offset [mas] & $\Delta \alpha \cos \delta= +98 \pm 10$ \\
& $\Delta \delta = -313 \pm 10$ \\    
North lobe integrated flux [mJy]  & $18.9 \pm 0.1$ \\
South lobe integrated flux [mJy]  & $17.2 \pm 0.1$ \\
Position angle P.A. [deg] & $163\pm 1$ \\
\hline
\end{tabular}
\end{table}

\section{Discussion and conclusions}

We assume that the northern and southern extensions of the core-subtracted radio map (Fig.~\ref{fig:blocks}, epoch 1999 September 16.037~UT) trace ejections from a symmetric bipolar jet launched by the central black hole. The morphology suggests a structure lying close to the plane of the sky. From the positions listed in Table~2 we obtain angular separations of $334 \pm 10$~mas (North) and $328 \pm 10$~mas (South). Adopting the ejection times proposed by \citet{Hjellming_2000} and the distance of $5.9 \pm 0.8$~kpc \citep{gaia2022}, the inferred apparent velocities would reach several tens of $c$, incompatible with a symmetric jet in the sky plane. However, considering earlier X-ray activity reported by \citet{Rusos2002}, the ejection could have occurred around September~1.5~UT, 
following a sharp drop
in flux observed in the lightcurve of BeppoSAX and RXTE/ASM
data, together with an extremely change in the corresponding Xray
spectrum. 
This would yield a more reliable subluminal apparent speeds of $\beta_{\rm a} = 0.8 \pm 0.1$ for both lobes.

The flux density ratio between condensations of the approaching and receding jet  at the same distance from the core ($S_a$ and $S_r$ respectively) is related to the Doppler boosting formula \citep{Pearson_1987} as:
\begin{equation}
    \frac{S_a}{S_r} = \left(\frac{1+\beta_j \cos i_j}{1-\beta_j \cos i_j}\right)^{k-\alpha},
\end{equation}
where $\beta_j$ is the intrinsic jet velocity in units of the speed of light $c$, $i_j$ is the ejection angle with respect to the line of sight, $\alpha$ is the radio spectral index of the jet
and $k = 2$ and $k = 3$ correspond to the continuous jet and the discrete blob case, respectively. 
On the other hand, the apparent speed of approaching ejecta $\beta_a$ is related to the intrinsic velocity $\beta_j$ of the jet by:
\begin{equation}
    \beta_a = \frac{\beta_j \sin i_j}{1- \beta_j \cos i_j}, 
\end{equation}
 For \bicho\ we know $\alpha = -0.75\pm0.05$ \citep{Hjellming_2000} and $S_a/S_r = 1.10\pm0.01$ (from Table~\ref{tab:prop}), so we can use these two last equations to derive both $\beta_j$ and $i_j$. As a result, $\beta_j = 0.8\pm 0.1$ independently of the continuity of the jet, while $i_j = 88.8^\circ\pm 0.1^\circ$ for a continuous jet and $i_j = 89.1^\circ\pm 0.1^\circ$ for the discrete blob case. The corresponding bulk Lorentz factor $\Gamma = (1-\beta_j^2)^{-1/2} = 1.7\pm0.4$.

 This inclination $i_j$ differs from the binary orbital value $i_{\rm o} = 72.3^{\circ} \pm 4.1^{\circ}$ \citep{MacDonald2014}, implying a misalignment angle $\theta_{\rm m} > 12.2^{\circ}\pm0.1^\circ$. 
 Such a spin--orbit offset could warp the disc through the Bardeen--Petterson effect \citep{BP75}, with misaligned outer regions inducing mild precession, as observed in SS~433. In support of this, a 
 Lomb–Scargle analysis of the available Zwicky Transient Facility (ZTF) data reveals a tentative $\sim$121-day modulation, broadly consistent with the precession range expected for a system with an orbital period of $2.82$ days \citep{Wijers99}.
On the other hand, the outflow has \(\Gamma \beta_j \geq 1\), placing it  in the regime of \textit{fast jets} according to \citet{Fender25}. These authors statistically infer that such jets are 
 likely launched near the spin axis through a Blandford--Znajek process, consistent with repeated ejections along a more stable axis without precession.
 Given the current observational uncertainties and the tentative nature of the ZTF signal, both scenarios remain viable, and we do not exclude either possibility at this stage.

In any case, 
the radio jet in \bicho\ would have a P.A.~$\approx 162^{\circ}$, remarkably aligned to 
the extended $>25$~TeV emission detected by LHAASO \citep[see the red line overlaid to the Test Statistic TS map in][]{lhaaso2025} and the map by H.E.S.S. \citep{HESS2025}. This  strongly suggests that the UHE particles are channeled along the same relativistic outflow traced in radio. Assuming $\Gamma \approx 1.7$ and a nearly sky-plane orientation, Doppler boosting is modest and the observed luminosity largely intrinsic. For a distance of $5.9$~kpc and peak X-ray fluxes of a few $\times 10^{-7}$~erg~cm$^{-2}$~s$^{-1}$ in the energy band 1--12 keV \citep{Rusos2002}, we estimate $L_{\rm X} \approx 10^{39}$~erg~s$^{-1}$, near the Eddington limit for a stellar-mass black-hole, similar (or not far) to the
corresponding value for SS~433.
On the other hand, the extended TeV-PeV emission has an observed flux $\lesssim 10^{-11}~\mathrm{erg~cm^{-2}~s^{-1}}$\citep{Alfaro_2024,lhaaso24}, 
 corresponding to a luminosity $L_\gamma \lesssim 4\times 10^{34} ~\mathrm{erg~s^{-1}}$, so the jets seen in radio seem capable to provide the required injection power.

A leptonic origin for the extended UHE emission faces severe quantitative challenges once the $\sim100$\,pc scale of the gamma-ray structure is considered. As emphasized by \citet{Alfaro_2024}, electrons energetic enough to generate $\sim200$\,TeV photons would require fast in--situ acceleration and must remain radiatively efficient over distances far exceeding their expected cooling length. For plausible lobe magnetic fields ($B\sim5$--$10\,\mu$G), the synchrotron cooling time of $E_e\gtrsim100$–200\,TeV electrons is only $\sim10^2$–$10^3$\,yr, far shorter than the diffusion time across $\sim100$\,pc, even under Bohm-like conditions ($t_{\rm diff}\gg10^3$\,yr). Moreover, inverse Compton scattering at these energies proceeds deep in the Klein–Nishina regime, which strongly suppresses the radiative efficiency and further increases the energetic burden of any leptonic model. The \citet{lhaaso24} reports a comparable difficulty, noting that IC emission cannot
reproduce the observed hard spectrum up to nearly 1~PeV unless an
unusually hard electron injection spectrum is assumed or the magnetic
field is extremely weak ($B \lesssim 0.5\,\mu$G).
Although the recent H.E.S.S.\ analysis \citep{HESS2025} shows that, under exceptionally low magnetic fields ($B\lesssim3\,\mu$G) and unusually rapid particle propagation, a leptonic contribution cannot be entirely excluded, 
hadronic interactions of relativistic protons remains the more natural explanation for both the hard spectrum and the large spatial extent of the UHE emission, given their longer cooling times and ability to propagate to large distances. If the emission arises from $\pi^0$ decay following $p$--$p$ collisions, even with a low efficiency $\eta_{\rm had} \approx 0.01$, the required proton power is
\begin{equation}
L_{\rm p} \approx \frac{L_{\gamma}}{\eta_{\rm had}} \sim 4 \times 10^{36}~{\rm erg~s^{-1}},
\end{equation}
well within the X-ray energy budget. This suggests that V4641~Sgr could act as a Galactic super-PeVatron, with acceleration sites in jet--ISM interaction zones or termination shocks. In
this context, a jet directed close to the plane of the sky may facilitate
detection of such structures, which would be severely foreshortened
or Doppler-deboosted in a microblazar geometry.

For such hadronic emission, jet interaction with the ambient medium is essential, though the local density remains uncertain.
Current molecular gas maps are too noisy to confirm or rule out the presence of sufficient molecular hydrogen in the region \citep{Alfaro_2024}. Nevertheless, independent studies suggest that the source is heavily obscured by local material, indicating that dense gas may be present along the line of sight \citep{Rusos02,Koljonen2020}. Previous modeling has shown that, for a typical interstellar density of \( n \sim 1\,\mathrm{cm^{-3}} \), the gamma-ray bubble would form at a distance of approximately $R\simeq 0.76(L_X/m_p n)^{1/5}\tau^{3/5} \simeq 110$~pc from the central source \citep{Alfaro_2024}, with $m_p$ the mass of the proton and $\tau\sim 1$ Myr the estimated age of the system \citep{Salvesen_2020}. A lower density estimate of \( n \sim 0.3\,\mathrm{cm^{-3}} \), as suggested by recent X-ray observations \citep{Suzuki2025}, would push this interaction region farther out to around 140~pc. Both distances are compatible with the observed UHE structure.
Upcoming high-resolution radio observations with MeerKAT may reveal these interaction sites, as recently observed in Cyg~X-1  \citep{MeerKAT_CygX1}, GRS~1758$-$258 \citep{MeerKAT_nosotros}, and GRS~1915+105 \citep{MeerKAT_GRS}.

In summary, we propose a self-consistent scenario in which V4641~Sgr hosts mildly relativistic ($\Gamma \approx 1.7$) symmetric radio jets lying near the sky plane and aligned with the UHE $\gamma$-ray axis. This framework explains the radio, X-ray, and TeV--PeV emission through energetically plausible hadronic processes and naturally accounts for the observed extended morphology and lack of Doppler shift in the OIR spectrum of \bicho\ \citep{Chaty2003}. Interaction between the jet and the surrounding
medium is likely required then to explain the extended emission, and
may become detectable with future high-sensitivity radio campaigns.
Future multi-wavelength observations will be crucial to confirm the acceleration sites and dominant emission mechanisms in this exceptional microquasar.

\section*{Acknowledgements}

 The authors acknowledge support from project PID2022-136828NB-C42 funded by the Spanish MCIN/AEI/10.13039/501100011033 and ``ERDF A way of making Europe". We also acknowledge funding for open access charge by Universidad de Ja\'en. 
The National Radio Astronomy Observatory is a facilitiy of the U.S. National Science Foundation operated under cooperative agreement by Associated Universities, Inc.
This work also made use of data from the European Space Agency (ESA) mission
 {\it Gaia} (\url{https://www.cosmos.esa.int/gaia}), processed by the {\it Gaia}
 Data Processing and Analysis Consortium (DPAC,
\url{https://www.cosmos.esa.int/web/gaia/dpac/consortium}). Funding for the DPAC
 has been provided by national institutions, in particular the institutions
 participating in the {\it Gaia} Multilateral Agreement. 
PLE wants to thank Jos\'e Miguel and Pedro Lino for their most delightful conversations and for offering  their support so selflessly.
%%%%%%%%%%%%%%%%%%%%%%%%%%%%%%%%%%%%%%%%%%%%%%%%%%
\section*{Data Availability}

The data used in the preparation of this Letter are public and can be accessed in their respective databases.



\bibliographystyle{mnras}
\bibliography{V4641Sgr-2} % if your bibtex file is called example.bib

%%%%%%%%%%%%%%%%%%%%%%%%%%%%%%%%%%%%%%%%%%%%%%%%%%
% Don't change these lines
\bsp	% typesetting comment
\label{lastpage}
\end{document}